\documentstyle[12pt]{article}
\title{Comment on SU(16) grand unification}
\author{L.\ Lavoura\\
\small Department of Physics, Carnegie-Mellon University, \\
\small Pittsburgh, Pennsylvania 15213, U.S.A.}
\begin{document}
\maketitle
\begin{abstract}
In a recent paper on SU(16) grand unification,
because of the presence of intermediate-energy gauge groups
containing products of U(1) factors
which are not orthogonal among themselves,
the renormalization-group treatment has a few small errors.
I correct it.
I emphasize that one should not switch gauge couplings
at the various thresholds.
It is easier,
and it avoids errors,
to use throughout the gauge couplings of the standard model,
and compute at each threshold,
in the usual way,
the extra contributions to the beta functions
from the extra non-decoupled fields.
I also point out that the SU(16) grand unification theory,
due to the large number of scalars present in it,
is not asymptotically free.
It becomes a strong-coupling theory
at energies only slightly larger than the unification scale.
\end{abstract}

\vspace{5mm}

This is a Comment on the paper of Ref.\ \cite{pal},
which is on the Grand Unified Theory (GUT) SU(16).
The points made in this Comment are however
also relevant to the GUT SU(15),
notably its analysis in the paper of Ref.\ \cite{su15},
and also other GUTs.
Moreover,
the first two paragraphs of this Comment are of a general nature,
and one does not need to have read any of the above papers
in order to understand them.

In the context of GUTs,
when intermediate-energy gauge groups
containing products of U(1) factors arise,
people usually use the set of U(1) factors
which is more intuitive (which has some simple physical interpretation).
Those U(1) factors are usually not,
however,
orthogonal relative to the active (non-decoupled) set of fields
at each energy scale.
Orthogonality is defined as follows:
two U(1) charges are orthogonal if the sum over all the active fields
of the product of their values,
weighted by the factors $ 11/3 $ for gauge-boson fields,
$ -2/3 $ for fermion fields,
and $ -1/3 $ for complex scalar fields,
vanishes.
This non-orthogonality leads \cite{erro} to errors
in the renormalization-group (RG) treatment of the theory.
More or less erroneous conclusions
may then be extracted from the RG analysis;
I have given \cite{erro} as examples two GUTs,
the first one based on the group SO(10),
where the error has led only
to small quantitative imprecisions in the results of the RG analysis,
the second one based on the group SU(8),
where the error has led to qualitatively false conclusions being
extracted from the RG analysis.
I find the same error in Ref.\ \cite{pal},
this time in the context of the GUT SU(16).
It is a much more complex case.
The breaking chain of SU(16) in Ref.\ \cite{pal}
has several intermediate-energy gauge groups,
three of which contain products of two U(1) factors.
Moreover,
the authors of Ref.\ \cite{pal} have,
at each threshold energy,
traded the gauge couplings of one intermediate-energy gauge group
by the ones of the next intermediate-energy gauge group,
which procedure involves a matching of the
gauge couplings at each threshold,
described in detail in Eqs.\ (2.11) to (2.29) of Ref.\ \cite{pal}.
This procedure,
although correct,
is complicated and unnecessary.
I emphasize here an argument at the end of Ref.\ \cite{erro}.
The beta functions of the gauge couplings of the standard model
can only change in response to more fields becoming active,
{\it i.e.},
non-decoupled,
at some energy in the middle of the RG evolution.
That change is always computed from the Dynkin indices
of the representations of newly active fields.
For the purpose of the RG evolution,
it is irrelevant whether the effective gauge group changes or not
at the threshold
({\it i.e.},
whether at the threshold some gauge bosons become active,
or only fermions and/or scalars become active).
One just needs to compute the color representations
and the weak-isospin representations
and the weak hypercharges of all the fields
which have their masses at each threshold energy,
and take them into account in the normal fashion
in the computation of the beta functions.
This procedure is transparent,
and it avoids both the pitfall of the intermediate-energy gauge groups
with products of U(1) factors,
and the complication of having to trade some gauge couplings for others
with the use of matching functions.

What was said above should be physically intuitive.
Its mathematical justification is in the fact that
all the gauge groups at each energy range,
including the U(1) factors,
must be orthogonal to each other,
relative to the set of fields active at the particular energy.
The sum over the active fields
of the product of the eigenvalues of any two generators
of the gauge group must always vanish.
Else,
the gauge bosons coupled to those generators mix by means of the one-loop
vacuum-polarization diagrams.
Then,
it is meaningless to talk about the one-loop RG evolution
of the gauge couplings of each of those gauge bosons,
because at the one-loop level those gauge bosons are mixing
during their propagation.
Now,
if all the generators of the intermediate-energy group are duly orthogonal,
then the argument at the end of Ref.\ \cite{erro} shows that
the following two procedures lead to the same result:
performing the RG evolution in terms of the intermediate-energy groups,
or performing all the RG evolution in terms of the gauge couplings
of the standard model.
The second procedure is much simpler.

In the breaking chain of SU(16) studied in Ref.\ \cite{pal}
there are three intermediate-energy gauge groups
containing factors U(1)$\otimes$U(1).
If we consider Eqs.\ (2.31) and (2.32) of Ref.\ \cite{pal},
we observe that the beta functions of the U(1) factors
are not needed in the computation of the evolution of $ \alpha_{3c} $
and of $ \alpha_{2L} $.
Therefore,
that computation is correct (I have checked it).
The problem is with the computation
of the evolution of the hypercharge gauge coupling.
Tables 1 and 2 give the set of scalar fields active
at each particular energy.
\begin{table}
\centering
\begin{tabular}{||c|c|c|c|c|c|}
\hline
\hline
  &
$ 16 $  &
$ 4^l 12^q $  &
$ 4^l 6^q_L 6^q_R 1^q_B $  &
$ 4^l 3^q_L 2^q_L 6^q_R 1^q_B $  &
$ 4^l 3^q_L 2^q_L 3^u_R 3^d_R 1^q_B 1^q_{\Lambda} $  \\
\hline
\hline
complex  &
$ 1820 $  &
$ (1, 1) $  &
  &  &  \\
\hline
real  &
$ 255_1 $  &
$ (1, 143) $  &
$ (1, 1, 1) [0] $  &
  &  \\
\hline
real  &
$ 14144 $  &
$ (1, 4212) $  &
$ (1, 189, 1) [0] $  &
$ (1, 1, 1, 1) [0] $  &
  \\
\hline
real  &
$ 255_2 $  &
$ (1, 143) $  &
$ (1, 1, 35) [0] $  &
$ (1, 1, 1, 35) [0] $  &
$ (1, 1, 1, 1, 1) [0, 0] $  \\
\hline
complex  &
$ 560_1 $  &
$ (1, 220) $  &
$ (1, 1, 20) [- \frac{3}{2 \sqrt{6}}] $  &
$ (1, 1, 1, 20) [- \frac{3}{2 \sqrt{6}}] $  &
$ (1, 1, 1, \bar{3}, 3) [- \frac{3}{2 \sqrt{6}}, - \frac{1}{2 \sqrt{3}}] $  \\
\hline
real  &
$ 255_3 $  &
$ (15, 1) $  &
$ (15, 1, 1) [0] $  &
$ (15, 1, 1, 1) [0] $  &
$ (15, 1, 1, 1, 1) [0, 0] $  \\
\hline
complex  &
$ 16 $  &
$ (4, 1) $  &
$ (4, 1, 1) [0] $  &
$ (4, 1, 1, 1) [0] $  &
$ (4, 1, 1, 1, 1) [0, 0] $  \\
\hline
complex  &
$ 560_2 $  &
$ (4, 66) $  &
$ (4, 6, 6) [0] $  &
$ (4, 3, 2, 6) [0] $  &
$ (4, 3, 2, 1, \bar{3}) [0, - \frac{1}{2 \sqrt{3}}] $  \\
\hline
complex  &
$ 136 $  &
$ (10, 1) $  &
$ (10, 1, 1) [0] $  &
$ (10, 1, 1, 1) [0] $  &
$ (10, 1, 1, 1, 1) [0, 0] $  \\
\hline
\hline
\end{tabular}
\caption[]{Scalar representations relevant for the symmetry breaking}
\end{table}
\begin{table}
\centering
\begin{tabular}{|c|c|c|c|c||}
\hline
\hline
$ 4^l 3^q_L 2^q_L 3^q_R 1^q_Y $  &
$ 3^l 3^q_L 2^q_L 3^q_R 1^l_X 1^q_Y $  &
$ 2^l_L 3^q_L 2^q_L 3^q_R 1^l_Y 1^q_Y $  &
$ 3_c 2_L 1_Y $  &
$ 3_c 1_Q $ \\
\hline
\hline
  &  &  &  &  \\
\hline
  &  &  &  &  \\
\hline
  &  &  &  &  \\
\hline
  &  &  &  &  \\
\hline
$ (1, 1, 1, 1) [0] $  &
  &  &  &  \\
\hline
$ (15, 1, 1, 1) [0] $  &
$ (1, 1, 1, 1) [0, 0] $  &
  &  &  \\
\hline
$ (4, 1, 1, 1) [0] $  &
$ (3, 1, 1, 1) [\frac{1}{2 \sqrt{6}}, 0] $  &
$ (1, 1, 1, 1) [0, 0] $  &
  &  \\
\hline
$ (4, 3, 2, \bar{3}) [\frac{3}{2 \sqrt{33}}] $  &
$ (3, 3, 2, \bar{3}) [\frac{1}{2 \sqrt{6}}, \frac{3}{2 \sqrt{33}}] $  &
$ (2, 3, 2, \bar{3}) [- \frac{1}{2 \sqrt{3}}, \frac{3}{2 \sqrt{33}}] $  &
$ (1, 1) [0] $  &
  \\
\hline
$ (10, 1, 1, 1) [0] $  &
$ (3, 1, 1, 1) [- \frac{2}{2 \sqrt{6}}, 0] $  &
$ (2, 1, 1, 1) [\frac{1}{2 \sqrt{3}}, 0] $  &
$ (1, 2) [\frac{1}{2} \sqrt{\frac{3}{5}}] $  &
$ (1) [0] $  \\
\hline
\hline
\end{tabular}
\caption[]{Continuation of Table 1}
\end{table}
It is clear that,
because of the representation $ 560_1 $,
the charges $ U(1)^q_B $ and $ U(1)^q_{\Lambda} $
in the intermediate-energy gauge group
$ 4^l 3^q_L 2^q_L 3^u_R 3^d_R 1^q_B 1^q_{\Lambda} $ are not orthogonal.
As a consequence,
the contribution in Ref.\ \cite{pal} of the $ 560_1 $
to the RG coefficient for the hypercharge-coupling RG running
between energies $ M_B $ and $ M_{6R} $
($9/10$) is wrong:
the correct value of that coefficient is 0,
as is evident from the fact that all the scalars of the $ 560_1 $
active at that energy scale have vanishing hypercharge.
Similarly,
the contribution of the $ 560_2 $ to the running of the hypercharge
used in Ref.\ \cite{pal}
is wrong in the energy range $ M_Y < M < M_{4l} $.
In the first part of that range,
$ M_{3l} < M < M_{4l} $,
the correct coefficient is $ 9/10 $ instead of $ 9/2 $,
and in the second part of that range,
$ M_Y < M < M_{3l} $,
the correct coefficient is 0 instead of $ 18/5 $
(once again,
because all the scalars of the $ 560_2 $ active at that energy scale
have zero hypercharge).

As a consequence of these small errors,
Eqs.\ (2.35) and (2.36) of Ref.\ [1] should be replaced by
\begin{eqnarray}
n_G & = &
\frac{2 \pi}{\ln 10} \left[
- \frac{457}{12096} \alpha_{3c}^{-1} (M_Z)
+ \frac{179}{1792} \alpha_{2L}^{-1} (M_Z)
- \frac{3005}{48384} \alpha_{1Y}^{-1} (M_Z)
\right]
- \frac{4943}{16128} n_Z
\nonumber\\
    &   &
+ \frac{397}{3024} n_Y - \frac{3439}{16128} n_{3l}
+ \frac{1171}{1728} n_{4l}
+ \frac{2285}{6048} n_B
+ \frac{47245}{16128} n_{6R} - \frac{41875}{16128} n_{6L}\, ,
\label{eq:nG}\\*[2mm]
n_{12} & = &
\frac{2 \pi}{\ln 10} \left[
- \frac{1}{24} \alpha_{3c}^{-1} (M_Z)
+ \frac{3}{32} \alpha_{2L}^{-1} (M_Z)
- \frac{5}{96} \alpha_{1Y}^{-1} (M_Z)
\right]
- \frac{7}{32} n_Z
\nonumber\\
       &   &
+ \frac{1}{6} n_Y - \frac{7}{32} n_{3l} + \frac{13}{24} n_{4l}
+ \frac{5}{12} n_B + \frac{85}{32} n_{6R} - \frac{75}{32} n_{6L}\, .
\label{eq:n12}
\end{eqnarray}
{}From these equations the correct versions of Eqs.\ (2.38) and (2.39),
and of Eqs.\ (2.41) and (2.42),
can easily be derived.
It may also be useful to give the value of $ \alpha_G (M_G) $:
\begin{eqnarray}
\alpha_G^{-1} (M_G) & = &
- \frac{37343}{72576} \alpha_{3c}^{-1} (M_Z)
+ \frac{5029}{10752} \alpha_{2L}^{-1} (M_Z)
+ \frac{86165}{290304} \alpha_{1Y}^{-1} (M_Z)
\nonumber\\*[1mm]
              &   &
+ \frac{\ln 10}{2 \pi} \left(
\frac{419735}{96768} n_Z
+ \frac{54035}{18144} n_Y
- \frac{218825}{96768} n_{3l}
- \frac{54523}{10368} n_{4l}
\right.
\nonumber\\*[1mm]
              &   &
\left.
+ \frac{186715}{36288} n_B
+ \frac{208283}{96768} n_{6R}
- \frac{686405}{96768} n_{6L}
\right)\, .
\label{eq:omegaG}
\end{eqnarray}
It is easy to check that these corrections to the results
in Ref.\ \cite{pal} may sometimes be highly relevant.
For instance,
for $ n_Y = 2.5 $,
$ n_{3l} = 2.6 $,
$ n_{4l} = 12.0 $,
$ n_B = 13.1 $,
$ n_{6R} = 13.5 $ and $ n_{6L} = 13.6 $,
Eqs.~\ref{eq:nG} and \ref{eq:n12} give
$ n_{12} = 13.71 $ and $ n_G = 13.81 $,
{\it i.e.},
$ M_{12} $ and $ M_G $ about two orders of magnitude smaller
than what would have been found from Ref.\ \cite{pal}.

A second comment that I want to make
is that the SU(16) theory has such an enormous number of scalars
that it is not asymptotically free and it is strongly coupled.
Let us calculate the precise extension of this effect.
The Dynkin indices $ l_R $
of the relevant representations $ R $ of SU(16) are
\begin{equation}
l_{16} = \frac{1}{2}\, ,\
l_{136} = 9\, ,\
l_{255} = 16\, ,\
l_{560} = \frac{91}{2}\, ,\
l_{1820} = 182\, ,\
l_{2160} = \frac{423}{2}\, ,\
l_{14144} = 1664\, .
\label{eq:Dynkin}
\end{equation}
The 2160 is contained in the product $ \overline{16} \times 136 $,
and I will refer to its use later.
The beta function for SU(16) is computed as in Eq.\ (2.4):
\begin{eqnarray}
\beta_{16} & = & \frac{11}{3} \times 16 - \frac{1}{3} \times 6
\nonumber\\
           &   &
- \frac{1}{6}
\left( 2 \times 182 + 16 + 1664 + 16 + 2 \times \frac{91}{2}
+ 16 + 2 \times \frac{1}{2} + 2 \times \frac{91}{2}
+ 2 \times 9 \right)
\nonumber\\
           & = & - \frac{1937}{6}\, .
\label{eq:beta16}
\end{eqnarray}
The beta function is negative,
which has as a consequence the existence of a Landau pole
[divergence of the gauge coupling of SU(16)].
The energy $ M_L $ at which that pole occurs
is given as a function of the unification energy $ M_G $ by
\begin{equation}
\ln \frac{M_L}{M_G} = \frac{12 \pi}{1937} \alpha^{-1}_G (M_G)\, .
\label{eq:Landau}
\end{equation}
For $ \alpha^{-1}_G (M_G) \sim 10 $,
one obtains $ M_L / M_G \sim 1.2 $.
This means that the SU(16) theory becomes strongly coupled
almost immediately after $ M_G $.
It is not clear to me what the cosmological consequences
of this fact might be.
But it is clear that the threshold effects at $ M_G $ are enormous
and that they are not calculable in perturbation theory.

In order to lessen the problem
(but not eliminate it),
one might break SU(12)$_q$ to
SU(3)$_c \otimes$ SU(2)$_{qL} \otimes$ U(1)$_{qY}$
directly by means of a complex 2160 of SU(16),
which contains a (1, 924) of SU(4)$_l \otimes$SU(12)$_q$.
This would avoid the 14144 of scalars,
which is the largest single responsible for the negativeness
of $ \beta_{16} $.
This is however not enough to render $ \beta_{16} $ positive.
It appears that,
if we want to have GUTs based on such large groups as SU(15) or SU(16),
we must accept that the GUT is non-perturbative for
practically the whole range of its validity,
from $ M_G $ up to the Planck energy.

\vspace{2mm}

I acknowledge useful discussions with Professors L.-F.\ Li
and L.\ Wolfenstein.
I also thank L.\ Wolfenstein for reading the manuscript.
This work was supported by the United States Department of Energy,
under the contract DE-FG02-91ER-40682.

\vspace{5mm}

%
%
\end{document}